\documentclass[spanish,english,a4paper,manuscript]{revtex4}
\usepackage[T1]{fontenc}
\usepackage[latin1]{inputenc}
\usepackage{amsmath}
\usepackage{amssymb}

\makeatletter
\@ifundefined{textcolor}{}
{%
 \definecolor{BLACK}{gray}{0}
 \definecolor{WHITE}{gray}{1}
 \definecolor{RED}{rgb}{1,0,0}
 \definecolor{GREEN}{rgb}{0,1,0}
 \definecolor{BLUE}{rgb}{0,0,1}
 \definecolor{CYAN}{cmyk}{1,0,0,0}
 \definecolor{MAGENTA}{cmyk}{0,1,0,0}
 \definecolor{YELLOW}{cmyk}{0,0,1,0}
 }

\usepackage{amstext}

\makeatletter
\@ifundefined{textcolor}{}
{%
 \definecolor{BLACK}{gray}{0}
 \definecolor{WHITE}{gray}{1}
 \definecolor{RED}{rgb}{1,0,0}
 \definecolor{GREEN}{rgb}{0,1,0}
 \definecolor{BLUE}{rgb}{0,0,1}
 \definecolor{CYAN}{cmyk}{1,0,0,0}
 \definecolor{MAGENTA}{cmyk}{0,1,0,0}
 \definecolor{YELLOW}{cmyk}{0,0,1,0}
 }

\makeatletter

\makeatletter

\makeatletter

\makeatletter

\makeatletter

\makeatletter

\usepackage{geometry}

\makeatother

\makeatother

\makeatother

\makeatother

\makeatother

\makeatother

\makeatother

\makeatother

\usepackage{babel}
\addto\shorthandsspanish{\spanishdeactivate{~<>}}

\begin{document}

\title{On the microscopic nature of dissipative effects in special relativistic
kinetic theory}

\author{A. L. Garcia-Perciante$^{1}$, A. Sandoval-Villalbazo$^{2}$, L.
S. Garcia-Colin$^{3}$}

\address{$^{1}$Depto. de Matematicas Aplicadas y Sistemas, Universidad Autonoma
Metropolitana-Cuajimalpa, Artificios 40 Mexico D.F 01120, Mexico.}

\address{$^{2}$Depto. de Fisica y Matematicas, Universidad Iberoamericana,
Prolongacion Paseo de la Reforma 880, Mexico D. F. 01219, Mexico.}

\address{$^{3}$Depto. de Fisica, Universidad Autonoma Metropolitana-Iztapalapa,
Av. Purisima y Michoacan S/N, Mexico D. F. 09340, Mexico. Also at
El Colegio Nacional, Luis Gonzalez Obregon 23, Centro Historico, Mexico
D. F. 06020, Mexico.\\
 }
\begin{abstract}
A microscopic formulation of the definition of both the heat flux
and the viscous stress tensor is proposed in the framework of kinetic
theory for relativistic gases emphasizing on the physical nature of
such fluxes. A Lorentz transformation is introduced as the link between
the laboratory and local comoving frames and thus between molecular
and chaotic velocities. With such transformation, the dissipative
effects can be identified as the averages of the chaotic kinetic energy
and the momentum flux out of equilibrium, respectively. Within this
framework, a kinetic foundation of the ensuing transport equations
for the relativistic gas is achieved. To our knowledge, this result
is completely novel.
\end{abstract}
\maketitle

\section{Introduction}

Relativistic kinetic theory is not a new subject, however it now finds
itself in a spotlight due to the increasing interest in relativistic
thermodynamics triggered by recent heavy ion collisions experiments,
electron-positron plasma generation, and the traditional astrophysical
applications of relativistic hydrodynamics. The theory has its roots
in original works by Jüttner \cite{juttner} for the equilibrium case
while the first kinetic theory treatment was formulated by Israel
\cite{Israel0-1}. In such work, the author finds an expression for
the stress energy tensor by solving the Boltzmann equation using a
Chapman-Enskog expansion. However, since the systematic (hydrodynamic)
and chaotic (or peculiar) components of the total velocity of a given
molecule are not explicitly distinguished, the different contributions
to this tensor cannot be identified in the same fashion as in the
non-relativistic case \cite{degrootNR,cc}. Instead, projections in
parallel and orthogonal directions with respect to the hydrodynamic
velocity of the stress energy tensor are used and the interpretations
of the different contributions agree with those that follow from the
phenomenological counterpart as developed by Eckart \cite{Eckart1-1}.
This procedure is essentially followed by most authors \cite{degroor,ck}.

On the other hand, in non-relativistic kinetic theory, a clear distinction
can be made between the effects caused by the {}``bulk'', or mechanical,
properties of the fluid and its microscopic ones. This permits the
identification of dissipative fluxes, i. e. heat and viscosity effects,
as averages of chaotic quantities \cite{cc}. In particular, the interpretation
of heat flux as the average of the chaotic kinetic energy flux, as
defined more than a century ago by R. Clausius \cite{clausius,brush}
and J. C. Maxwell \cite{MAX}, is asserted. This concept is absent
in the relativistic case, as was clearly noted in Ref. \cite{pa2001}.
In that work, a first proposal of how to introduce the chaotic velocity
concept in relativistic kinetic theory was put forward. In this work,
we follow the same line of thought and take it a step forward by explicitly
introducing Lorentz transformations in the stress-energy tensor integral
in order to separate mechanical and chaotic effects. By doing so,
we are able to clearly define the heat flux and the viscous stress
tensor as the average of chaotic energy and momentum fluxes in an
arbitrary frame, respectively.

To accomplish this task we have divided this work as follows. In Section
II, we briefly review the non-relativistic setup for calculating the
dissipative fluxes. The relativistic framework is introduced in Section
III where Lorentz transformations are used in order to introduce the
chaotic velocity and obtain the corresponding expressions for the
heat flux and viscous stress tensor. A brief discussion of the results
and final remarks are included in Section IV.

\section{non-relativistic kinetic theory}

Kinetic theory serves as the microscopic foundation of irreversible
thermodynamics and is capable of producing both the system of transport
equations as well as the constitutive equations needed in order to
make it a complete set describing the dynamics of fluids \cite{cc,degrootNR}.
As usual, the distribution function $f\left(\vec{r},\vec{v},t\right)$
is such that $f\left(\vec{r},\vec{v},t\right)d\vec{r}d\vec{v}$ is
the number of molecules contained in a 6-box in the phase space corresponding
to position $\vec{r}$ and molecular velocity $\vec{v}$. The local
variables are thus defined as averages weighted by this function.
The local particle density, hydrodynamic velocity and energy density
are thus defined as

\begin{equation}
n=\int f^{\left(0\right)}d^{3}v\label{eq:1}\end{equation}
\begin{equation}
\vec{u}=\frac{1}{n}\int\vec{v}f^{\left(0\right)}d^{3}v\label{eq:2}\end{equation}
\begin{equation}
e=\frac{1}{n}\int\frac{1}{2}mv^{2}f^{\left(0\right)}d^{3}v\label{eq:3-2}\end{equation}
respectively, where $f^{\left(0\right)}$ is the local equilibrium
distribution function:\begin{equation}
f^{\left(0\right)}\left(\vec{r},\vec{v},t\right)=n\left(\frac{m}{2\pi k_{B}T}\right)^{3/2}\exp\left(-\frac{m\left(\vec{v}-\vec{u}\right)^{2}}{2k_{B}T}\right)\label{eq:4-1}\end{equation}
$T$ being the temperature, $m$ the molecular mass and $k_{B}$ the
Boltzmann constant. The evolution of the distribution function is
given by the Boltzmann equation. For a simple (one component), non-degenerate,
diluted gas in the absence of external fields the kinetic equation
reads\begin{equation}
\frac{df}{dt}=J\left(ff'\right)\label{eq:5-1}\end{equation}
where, if $g$ and $\sigma$ are the relative velocity and cross section
for a collision between two particles respectively, the collision
term is given by\begin{equation}
J\left(f,\, f'\right)=\int\int\left\{ f\,'f_{1}\,'-f\, f_{1}\right\} g\sigma d\Omega dv_{1}^{3}\label{eq:6}\end{equation}
Primes denote quantities after the interaction and $\Omega$ is the
solid angle. The well known Maxwell-Boltzmann distribution function
given in Eq. (\ref{eq:4-1}) is precisely the solution of $J\left(f,\, f'\right)=0$,
namely the homogeneous Boltzmann equation. The solution to the inhomogeneous,
out of equilibrium, case is in general obtained via the Chapman-Enskog
method in which the general solution is written as\begin{equation}
f=f^{\left(0\right)}+f^{\left(1\right)}\label{eq:7-1}\end{equation}
where the second term contains corrections to the equilibrium solution
to first order in the gradients of the local variables and gives rise
to the dissipative fluxes. This term includes only dissipative effects
once the solubility constraints\begin{equation}
\int f^{\left(1\right)}d^{3}v=\int\vec{v}f^{\left(1\right)}d^{3}v=\int v^{2}f^{\left(1\right)}d^{3}v=0\label{eq:8}\end{equation}
are introduced such that the local variables are defined through the
local equilibrium distribution solely.

In this framework, the transport equations are obtained by multiplying
Eq. (\ref{eq:5-1}) by a collision invariant and integrating over
$\vec{v}$. Such procedure yields the Maxwell-Enskog transport equation\begin{equation}
\frac{\partial}{\partial t}\int\psi fd^{3}v+\nabla\cdot\int\psi\vec{v}fd^{3}v=0\label{eq:9}\end{equation}
which accounts for particle, momentum and energy balances for $\psi=1,\,\vec{v},\, v^{2}$,
respectively. Indeed, taking $\psi=1$ in Eq. (\ref{eq:7-1}) yields
the continuity equation\begin{equation}
\frac{\partial n}{\partial t}+\nabla\cdot\left(n\vec{u}\right)=0\label{eq:10}\end{equation}
For $\psi=\vec{v}$ one obtains\begin{equation}
\frac{\partial\left(n\vec{u}\right)}{\partial t}+\nabla\cdot\overleftrightarrow{\mathcal{T}}=0\label{eq:11}\end{equation}
where we introduced the stress tensor\begin{equation}
\overleftrightarrow{\mathcal{T}}=\int\vec{v}\vec{v}fd^{3}v\label{eq:12}\end{equation}
Finally, the energy balance is obtained for $\psi=v^{2}$:\begin{equation}
\frac{\partial ne}{\partial t}+\nabla\cdot\vec{J}_{e}=0\label{eq:13}\end{equation}
where we have defined the total energy flux as $\vec{J}_{e}=\int v^{2}\vec{v}fd^{3}v$. 

In order to isolate the purely dissipative contributions in $\overleftrightarrow{\mathcal{T}}$
and $\vec{J}_{e}$, one decomposes the molecular velocity in its two
basic components, usually written as \begin{equation}
\vec{v}=\vec{u}+\vec{k}\label{eq:df}\end{equation}
where $\vec{k}$ is the chaotic or peculiar component. In this case
such expression arises in a very natural way by observing the argument
of the exponential function in Eq. (\ref{eq:4-1}). It is clear that
$\int\vec{k}fd^{3}v=0$ in view of Eqs. (\ref{eq:4-1}) and (\ref{eq:10})
and thus\begin{equation}
e=\frac{1}{2}u^{2}+\varepsilon\label{eq:14}\end{equation}
\begin{equation}
\overleftrightarrow{\mathcal{T}}=n\vec{u}\vec{u}+nk_{B}T\mathbb{I}+\overleftrightarrow{\tau}\label{eq:15}\end{equation}
\begin{equation}
\vec{J}_{e}=\frac{1}{2}nu^{2}\vec{u}+n\vec{u}\varepsilon+n\vec{u}\cdot\left(nk_{B}T\mathbb{I}+\overleftrightarrow{\tau}\right)+\vec{q}\label{eq:16}\end{equation}
where\begin{equation}
n\varepsilon=\int\frac{k^{2}}{2}f^{\left(0\right)}d^{3}k=\frac{3}{2}k_{B}T\label{eq:17}\end{equation}
is the internal energy density per particle and the dissipative fluxes
are given by\begin{equation}
\overleftrightarrow{\tau}=\int\vec{k}\vec{k}f^{\left(1\right)}d^{3}k\label{eq:18}\end{equation}
\begin{equation}
\vec{q}=\int\frac{k^{2}}{2}\vec{k}f^{\left(1\right)}d^{3}k\label{eq:19}\end{equation}
Introducing these definitions in the transport equations and using
the local equilibrium assumption, one obtains the well known set of
hydrodynamic equations for the non-relativistic fluid.

In the next section, it will be shown how these ideas can be extrapolated
in a very natural way to the relativistic framework. In order to make
the transition more clear we want to point out at this stage the key
role of the transformation $\vec{v}=\vec{u}+\vec{k}$ in the formalism.
Notice that such a transformation can also be expressed in terms of
a Galilean matrix in space-time, that is\begin{equation}
v^{\mu}=\mathcal{G}^{\mu\nu}k_{\nu}\label{eq:23}\end{equation}
where the Galilean transformation is given by\begin{equation}
\mathcal{G}^{\mu\nu}=\left(\begin{array}{cccc}
1 & 0 & 0 & u_{x}/c\\
0 & 1 & 0 & u_{y}/c\\
0 & 0 & 1 & u_{z}/c\\
0 & 0 & 0 & 1\end{array}\right)\label{eq:24}\end{equation}
and\begin{equation}
v^{\mu}=\left(\begin{array}{c}
v_{x}\\
v_{y}\\
v_{z}\\
c\end{array}\right)\qquad k^{\mu}=\left(\begin{array}{c}
k_{x}\\
k_{y}\\
k_{z}\\
c\end{array}\right)\label{eq:25}\end{equation}
Whence, the decomposition $\vec{v}=\vec{u}+\vec{k}$ can be viewed
as a change in reference frames where an observer comoving with the
volume element of the fluid whose hydrodynamic velocity is $\vec{u}$
will measure a given molecule's velocity as $\vec{k}$ while an observer
in the laboratory sees the molecule moving at velocity $\vec{v}$
as given by Eq. (\ref{eq:23}).

\section{Relativistic kinetic theory}

In this section we will address the properties of a dilute, neutral,
non-degenerate gas within the realms of special relativity. This system
is thus described in a Minkowsky space-time whose metric is given
by $ds^{2}=dx^{2}+dy^{2}+dz^{2}-cdt^{2}$. For the molecules in this
gas, the molecular four-velocity is given by\begin{equation}
v^{\mu}=\gamma_{w}\left(\vec{w},c\right)\label{eq:26}\end{equation}
where\begin{equation}
\gamma_{w}\equiv\gamma\left(w\right)=\left(1-\frac{w^{2}}{c^{2}}\right)^{-1/2}\label{eq:k}\end{equation}
and $\vec{w}$ is the velocity. The distribution function has the
same interpretation as above, being $f\left(x^{\nu},v^{\nu}\right)d^{3}xd^{3}v$
the occupation number of a phase space cell. The special relativistic
Boltzmann equation in the absence of external forces is given by\begin{equation}
v^{\alpha}f_{,\alpha}=\dot{f}=J(ff')\label{eq:27}\end{equation}
where the collision term is defined as \begin{equation}
J\left(f,\, f'\right)=\int\int\left\{ f\,'f_{1}\,'-f\, f_{1}\right\} F\sigma d\Omega dv_{1}^{*}\label{eq:28}\end{equation}
Here $F$ is an invariant particle flux \cite{ck} which plays the
role of the relative velocity, $\sigma$ is the collision cross section
and the invariant differential volume in velocity space is $dv^{*}=\frac{cd^{3}v}{v^{4}}$.

Here, as in our previous works, the proposed solution method for the
kinetic equation is the Chapman-Enskog procedure to first order in
the gradients \cite{Nos1}. As has been shown elsewhere \cite{pa2009},
this solution leads to a constitutive equation for the heat flux in
terms of gradients of the state variables. This is consistent with
Onsager's regression of fluctuations hypothesis and thus predicts
no pathological behaviors in the system of hydrodynamic equations
\cite{jnnfm}. Additionally, this system of equations to first order
in the gradients, as predicted by kinetic theory, has been shown to
present no causality issues both in the non-relativistic and relativistic
cases \cite{jnnfm}. In this case, the local equilibrium function
is the Jüttner function\begin{equation}
f^{(0)}=\frac{n}{4\pi c^{3}z\mathcal{K}_{2}\left(\frac{1}{z}\right)}\exp\left(\frac{\mathcal{U}^{\beta}v_{\beta}}{zc^{2}}\right)\label{eq:29}\end{equation}
where $\mathcal{U}^{\beta}=\gamma_{u}\left(\vec{u},c\right)$ is the
hydrodynamic four-velocity, $z=k_{B}T/mc^{2}$ is the relativistic
parameter and $\mathcal{K}_{n}$ is the n-th order modified Bessel
function of the second kind. As in the non-relativistic case, the
transport equations are obtained by multiplying Eq. (\ref{eq:27})
by collision invariants, in this case $\psi=1,\, v^{\mu}$. Indeed,
the corresponding transport equation in the absence of external forces
is\begin{equation}
\left[\int v^{\alpha}\psi fdv^{*}\right]_{;\alpha}=0\label{eq:30}\end{equation}
which yields the continuity equation for $\psi=1$, the energy-momentum
balance equation for $\psi=mv^{\beta}$; that is, the momentum balance
in the absence of external forces for $\beta=1,2,3$ and the energy
balance for $\beta=4$. Equation (\ref{eq:30}) can be expressed in
a more conventional form as a general conservation law for four-flows
by defining the particle and stress-energy fluxes as\begin{equation}
N^{\nu}=\int v^{\nu}fdv^{*}\label{eq:31}\end{equation}
\begin{equation}
T^{\mu\nu}=m\int v^{\mu}v^{\nu}fdv^{*}\label{eq:32}\end{equation}
respectively. Thus, the transport equations are given by $N_{;\nu}^{\nu}=0$
and $T_{;\nu}^{\mu\nu}=0$. It is then appealing to write the integrals
in Eqs. (\ref{eq:31}) and (\ref{eq:32}) in terms of systematic and
chaotic quantities in order to separate the different contribution
to the fluxes, as done in the previous section (see Eqs. (\ref{eq:14})
to (\ref{eq:19})). To accomplish this, an appropriate transformation
law has to be assigned in order to introduce the chaotic velocity.
It is important to recall at this point that the hydrodynamic velocity
is a local equilibrium quantity and is thus only defined in each differential
volume where local equilibrium is assumed. If we fix our attention
in a single random molecule, we can consider two reference frames,
one in the laboratory ($S$) and one fixed in the volume where the
molecule is contained. This second frame ($\bar{S}$), in which the
molecules would be seen static on the average, is moving with a speed
$\vec{u}$ as seen by an observer fixed in $S$. Thus, observers in
$\bar{S}$ and $S$ would report that the corresponding velocities
are given by\begin{equation}
\bar{v}^{\alpha}=\gamma_{k}\left(\vec{k},c\right)\label{eq:33}\end{equation}
and\begin{equation}
v^{\beta}=\mathcal{L}_{\alpha}^{\beta}\bar{v}^{\alpha}=\mathcal{L}_{\alpha}^{\beta}K^{\alpha}\label{eq:34}\end{equation}
respectively. Here $\mathcal{L}_{\alpha}^{\beta}$ is a Lorentz boost
with velocity $\vec{u}$ and $K^{\alpha}=\gamma_{k}\left(\vec{k},c\right)$
is the chaotic four-velocity \cite{pa2001}. We wish to remind the
reader at this point that the contravariant transformation given in
Eq. (\ref{eq:34}) is equivalent to the relativistic velocity addition
law.

With the transformation given in Eq. (\ref{eq:34}), Eqs. (\ref{eq:31})
and (\ref{eq:32}) can be written as\begin{equation}
N^{\mu}=\mathcal{L}_{\alpha}^{\mu}\int K^{\alpha}fdK^{*}\label{eq:35}\end{equation}
\begin{equation}
T^{\mu\nu}=m\mathcal{L}_{\alpha}^{\mu}\mathcal{L}_{\beta}^{\nu}\int K^{\alpha}K^{\beta}fdK^{*}\label{eq:36}\end{equation}
where use has been made of the fact that, since $dv^{*}$ is an invariant
quantity, $dv^{*}=dK^{*}$. Also the equilibrium distribution function
given in Eq. (\ref{eq:29}) can be written in terms of the chaotic
speed by use of the invariant $\gamma_{k}=\mathcal{U}^{\beta}v_{\beta}/c^{2}$
in a similar fashion as in the the non relativistic case where the
argument of the Maxwellian is proportional to $k^{2}$. These two
properties which allow the calculation of integrals in terms of $K$
are verified in the Appendix. In order to obtain a general expression
for $T^{\mu\nu}$, we introduce an irreducible decomposition relative
to the hydrodynamic four-velocity direction. That is, in this 3+1
representation a second rank tensor can be expressed as \cite{Eckart1-1}\begin{equation}
T^{\mu\nu}=\tau\mathcal{U}^{\mu}\mathcal{U}^{\nu}+\tau^{\mu}\mathcal{U}^{\nu}+\tau^{\nu}\mathcal{U}^{\mu}+\tau^{\mu\nu}\label{eq:37}\end{equation}
where $\tau^{\mu}\mathcal{U}_{\mu}=0$ and $\tau^{\mu\nu}\mathcal{U}_{\nu}=0$.
The scalar, first and second rank tensors introduced can be expressed
in terms of $T^{\mu\nu}$ as \begin{equation}
\tau=T^{\mu\nu}\frac{\mathcal{U}_{\mu}\mathcal{U}_{\nu}}{c^{4}}\label{eq:38}\end{equation}
\begin{equation}
\tau^{\mu}=-\frac{1}{c^{2}}h_{\alpha}^{\mu}T^{\alpha\beta}\mathcal{U}_{\beta}\label{eq:39}\end{equation}
\begin{equation}
\tau^{\mu\nu}=h_{\alpha}^{\mu}h_{\beta}^{\nu}T^{\alpha\beta}\label{eq:40}\end{equation}
respectively. Here $h^{\mu\nu}=g^{\mu\nu}+\mathcal{U}^{\mu}\mathcal{U}^{\nu}/c^{2}$
is the well known projector which satisfies $\mathcal{U}_{\mu}h_{\nu}^{\mu}=0$.
It is important to point out in this stage that in the phenomenological
treatment, the quantities above are identified as the internal energy,
heat flux and stress tensor without a kinetic theory based justification
\cite{Eckart1-1}. These definitions are in turn used in most kinetic
treatments \cite{Israel0-1,degroor,cc}. It is precisely the aim of
this work to deduce, from purely kinetic grounds, that these quantities
are indeed related to internal energy, heat flux and stress \emph{interpreted
as averages over chaotic velocities} in a similar fashion as in Eqs.
(\ref{eq:17})-(\ref{eq:19}). 

The scalar $\tau$ can be calculated as\begin{equation}
\tau=m\frac{\mathcal{U}_{\mu}\mathcal{U}_{\nu}}{c^{4}}\int v^{\mu}v^{\nu}fdv^{*}=m\int\gamma_{k}^{2}fdK^{*}\label{eq:41}\end{equation}
which is the internal energy per particle. To see that this is so,
consider Eq. (\ref{eq:30}) with $\psi=mv^{4}$\begin{equation}
\frac{\partial}{\partial t}\left(m\int v^{4}v^{4}fdv^{*}\right)+\frac{\partial}{\partial x^{\ell}}\left(m\int v^{4}v^{\ell}fdv^{*}\right)=0\label{eq:42}\end{equation}
where here, as in the rest of this work, latin indices run from 1
to 3 only. It is clear from Eq. (\ref{eq:42}) that the integral in
the first term is indeed the total energy while the second integral
is the energy flux. Thus, the equivalent to the total energy moment
calculated in a rest frame yields the internal energy only, that is\begin{equation}
n\varepsilon=mc^{2}\int\gamma_{k}^{2}fdK^{*}\label{eq:43}\end{equation}
and thus,\begin{equation}
\tau=\frac{n\varepsilon}{c^{2}}=nm\left(3z+\frac{\mathcal{K}_{3}\left(\frac{1}{z}\right)}{\mathcal{K}_{2}\left(\frac{1}{z}\right)}\right)\label{eq:44}\end{equation}
For the vector quantity $\tau^{\mu}$ we have\begin{equation}
\tau^{\mu}=-\frac{1}{c^{2}}h_{\alpha}^{\mu}\mathcal{U}_{\beta}\int v^{\alpha}v^{\beta}fdv^{*}\label{eq:45}\end{equation}
which, using again the fact that $\mathcal{U}_{\beta}v^{\beta}=-c^{2}\gamma_{k}$
can be expressed as an integral over the chaotic velocities as follows\begin{equation}
\tau^{\mu}=h_{\alpha}^{\mu}\mathcal{L}_{\beta}^{\alpha}\int\gamma_{k}K^{\beta}fdK^{*}\label{eq:46}\end{equation}
It can be shown (see the Appendix) that the contraction of the projector
with the Lorentz transformation yields a tensor $\mathcal{R}_{\nu}^{\mu}=h_{\alpha}^{\mu}\mathcal{L}_{\nu}^{\alpha}$
given by\begin{equation}
\mathcal{R}_{4}^{\mu}=0\label{eq:47}\end{equation}
\begin{equation}
\mathcal{R}_{a}^{\mu}=\mathcal{L}_{a}^{\mu}\quad\text{for }a=1,\,2,\,3\label{eq:48}\end{equation}
and thus\begin{equation}
\tau^{\mu}=\mathcal{R}_{\beta}^{\mu}\int\gamma_{k}K^{\beta}fdK^{*}\label{eq:49}\end{equation}
We now introduce the Chapman-Enskog expansion\begin{equation}
\tau^{\mu}=\mathcal{R}_{\beta}^{\mu}\int\gamma_{k}K^{\beta}f^{\left(0\right)}dK^{*}+\mathcal{R}_{\beta}^{\mu}\int\gamma_{k}K^{\beta}f^{\left(1\right)}dK^{*}\label{eq:50}\end{equation}
and notice that the first terms vanishes since, for $\beta=1,2,3$
the integral $\mathcal{R}_{\beta}^{\mu}\int\gamma_{k}K^{\beta}f^{\left(0\right)}dK^{*}$
is odd in $k$ and the $\beta=4$ term in the sum is zero because
$\mathcal{R}_{4}^{\mu}=0$ for any $\mu$. Thus, only the integral
with $f^{\left(1\right)}$ survives and we can write\begin{equation}
\tau^{\mu}=\mathcal{R}_{\beta}^{\mu}\int\gamma_{k}K^{\beta}f^{\left(1\right)}dK^{*}\label{eq:51}\end{equation}
In order to re-introduce the Lorentz transformation, we notice that
\begin{equation}
\int\gamma_{k}K^{4}f^{\left(1\right)}dK^{*}=0\label{eq:53}\end{equation}
since the internal energy, as all state variables, is obtained only
through the equilibrium solution. That is, the subsiadiary condition,
which the solution $f^{\left(1\right)}$ will be enforced to satisfy,
requires\begin{equation}
\int\gamma_{k}^{2}f^{\left(i\right)}dK^{*}=0\qquad\text{for }i\neq0\label{eq:54}\end{equation}
Using Eq. (\ref{eq:53}), one can write Eq. (\ref{eq:51}) back in
terms of $\mathcal{L}_{\beta}^{\mu}$ which yields \begin{equation}
\tau^{\mu}=\mathcal{L}_{\beta}^{\mu}\int\gamma_{k}K^{\beta}f^{\left(1\right)}dK^{*}\label{eq:55}\end{equation}
By inspection of Eq. (\ref{eq:42}) one concludes that the integral
$\int\gamma_{k}K^{b}f^{\left(1\right)}dK^{*}$ is the heat flux in
a rest frame where $v^{\alpha}=K^{\alpha}$, and thus \begin{equation}
q_{\left[0\right]}^{\beta}=c^{2}\int\gamma_{k}K^{\beta}f^{\left(1\right)}dK^{*}\label{eq:58}\end{equation}
This expression is analogous to the one found in the non-relativistic
case and full of physical content. The heat flux is physically the
average flux of the chaotic energy, and Eq. (\ref{eq:58}) is completely
consistent which this idea. Now, in an arbitrary frame\begin{equation}
\tau^{\mu}=\frac{1}{c^{2}}\mathcal{L}_{\nu}^{\mu}q_{\left[0\right]}^{\nu}\label{eq:59}\end{equation}
which is, to the authors' knowledge, the first time that the heat
flux is obtained only from a kinetic theory standpoint as the average
of the peculiar kinetic energy flux of the molecules.

For the second rank tensor in the stress-energy tensor decomposition,
we calculate from Eq. (\ref{eq:40})\begin{equation}
\tau^{\mu\nu}=mh_{\alpha}^{\mu}h_{\beta}^{\nu}\mathcal{L}_{\eta}^{\alpha}\mathcal{L}_{\delta}^{\beta}\int K^{\eta}K^{\delta}fdK^{*}\label{eq:60}\end{equation}
or\begin{equation}
\tau^{\mu\nu}=m\mathcal{R}_{\eta}^{\mu}\mathcal{R}_{\delta}^{\nu}\int K^{\eta}K^{\delta}\left(f^{\left(0\right)}+f^{\left(1\right)}\right)dK^{*}\label{eq:61}\end{equation}
For the local-equilibrium term we have\begin{equation}
m\mathcal{R}_{\eta}^{\mu}\mathcal{R}_{\delta}^{\nu}\int K^{\eta}K^{\delta}f^{\left(0\right)}dK^{*}=m\mathcal{R}_{a}^{\mu}\mathcal{R}_{b}^{\nu}\int K^{a}K^{b}f^{\left(0\right)}dK^{*}\label{eq:62}\end{equation}
Since $f^{\left(0\right)}$ is even in $k$, only the $a=b$ terms
survive and thus\begin{equation}
m\mathcal{R}_{\eta}^{\mu}\mathcal{R}_{\delta}^{\nu}\int K^{\eta}K^{\delta}f^{\left(0\right)}dK^{*}=ph^{\mu\nu}\label{eq:63}\end{equation}
where we have introduced the well known result for the hydrostatic
pressure\begin{equation}
p=m\int\left(K^{1}\right)^{2}f^{\left(0\right)}dK^{*}=m\int\left(K^{2}\right)^{2}f^{\left(0\right)}dK^{*}=m\int\left(K^{3}\right)^{2}f^{\left(0\right)}dK^{*}\label{eq:64}\end{equation}
and\begin{equation}
p=nk_{B}T\label{eq:gg}\end{equation}
together with the identity\begin{equation}
\mathcal{R}_{a}^{\mu}\mathcal{R}_{b}^{\nu}\delta^{ab}=\mathcal{L}_{a}^{\mu}\mathcal{L}_{b}^{\nu}\delta^{ab}=h^{\mu\nu}\label{eq:65}\end{equation}
Equation (\ref{eq:63}) was obtained in a similar fashion (using Lorentz
transformations) by Weinberg \cite{weinberg}, nevertheless he did
not address the dissipative case following a kinetic theory approach. 

For the dissipative term in Eq. (\ref{eq:61}), which we write as
$\Pi^{\mu\nu}$, we have \begin{equation}
\Pi^{\mu\nu}=m\mathcal{R}_{\eta}^{\mu}\mathcal{R}_{\delta}^{\nu}\int K^{\eta}K^{\delta}f^{\left(1\right)}dK^{*}=m\mathcal{L}_{a}^{\mu}\mathcal{L}_{b}^{\nu}\int K^{a}K^{b}f^{\left(1\right)}dK^{*}\label{eq:66}\end{equation}
If $\Pi_{\left[0\right]}^{\alpha\beta}$ is the Navier-Newton tensor
calculated in a frame where the fluid is at rest\[
\Pi_{\left[0\right]}^{\mu\nu}=mh_{\alpha}^{\mu}h_{\beta}^{\nu}\int K^{\alpha}K^{\beta}fdK^{*}=m\delta_{a}^{\mu}\delta_{b}^{\nu}\int K^{a}K^{b}fdK^{*}\]
since in such frame $h_{4}^{\mu}=0$ and $h_{a}^{\mu}=\delta_{a}^{\mu}$.
Thus, the second rank tensor introduced in the stress-energy tensor
is\[
\tau^{\mu\nu}=ph^{\mu\nu}+\Pi^{\mu\nu}\]
where \begin{equation}
\Pi^{\mu\nu}=\mathcal{L}_{\alpha}^{\mu}\mathcal{L}_{\beta}^{\nu}\Pi_{\left[0\right]}^{\alpha\beta}\label{eq:333}\end{equation}
is the Navier-Newton tensor in an arbitrary frame.

\section{Summary and final remarks}

In the previous section, the different contributions to the stress-energy
tensor for a single component, dilute gas in the framework of special
relativity have been calculated by separating hydrodynamic and chaotic
contributions to the molecular velocities. This has been accomplished
by introducing Lorentz transformations to relate the velocity of a
molecule as measured by an arbitrary observer with the one measured
within a differential volume moving at the corresponding hydrodynamic
velocity, an idea introduced by two of us in Ref. \cite{pa2001},
combined with Eckart's decomposition \cite{Eckart1-1}.

The main results of this work can be summarized in the fact that all
quantities appearing in Eq. (\ref{eq:37}) have been obtained strictly
from kinetic theory \emph{using the concept of chaotic velocity}.
The first two terms are the equilibrium parts of the stress-energy
tensor and are well known. The main accomplishment of the calculation
here shown are the dissipative terms which appear here in a natural
way as averages over kinetic energy and momentum fluxes once the transformation
between molecular and peculiar velocities is introduced. Also it has
been shown that the heat flux transforms as a first rank tensor.

A kinetic derivation of the stress-energy tensor for a dissipative
fluid from first principles in kinetic theory has been lacking for
some time and thus hindering a clear derivation of the relativistic
Navier-Stokes equations. Equations (\ref{eq:35}) and (\ref{eq:36})
satisfy both needs and, in turn, pose a new question. Since both heat
and momentum fluxes in Eq. (\ref{eq:37}) are given by Eqs. (\ref{eq:59})
and (\ref{eq:333}) respectively, the hydrodynamic velocity factors
introduced by the Lorentz transformations will induce new non-linearities
in the system of hydrodynamic equations. This could yield new relativistic
effects for the relativistic gas which may be measurable. This question
and will be addressed in the future.

\section*{Appendix}

In this appendix the relations\begin{equation}
dv^{*}=dK^{*}\label{eq:70}\end{equation}
and\begin{equation}
\mathcal{U}^{\nu}v_{\nu}=\gamma_{k}\label{eq:71}\end{equation}
are shown to hold where $\mathcal{U}^{\mu}$, $v^{\mu}$ and $K^{\mu}$
are the hydrodynamic, molecular and chaotic four-velocities respectively.
Also, we verify that the tensor quantity $\mathcal{R}_{\nu}^{\mu}$
is indeed given by Eqs. (\ref{eq:47}) and (\ref{eq:48}).

To verify Eqs. (\ref{eq:70}) and (\ref{eq:71}), we consider two
reference frames $S$ and $\bar{S}$ with a relative speed $\vec{u}$
with respect to each other. That is, $S$ may be considered the laboratory
frame while $\bar{S}$ is a frame fixed to a volume element in the
fluid. For the sake of simplicity, we take the $x$ direction parallel
to $\vec{u}$. In this situation, we have three four-vectors related
to a given molecule in such fluid element\[
K^{\nu}=\gamma_{k}\left(\vec{k},c\right)\qquad\text{velocity of the molecule as measured by an observer in }\bar{S}\]
\[
v^{\nu}=\gamma_{v}\left(\vec{w},c\right)\qquad\text{velocity of the molecule as measured by an observer in }S\]
\[
\mathcal{U}^{\nu}=\gamma_{u}\left(u,0,0,c\right)\qquad\text{relative velocity between }S\text{ and }\bar{S}\]
The relationship between tensors in both references frames given by
the Lorentz transformation\begin{equation}
\mathcal{L}_{\nu}^{\mu}=\left[\begin{array}{cccc}
\gamma_{u} & 0 & 0 & \frac{u}{c}\gamma_{u}\\
0 & 1 & 0 & 0\\
\frac{u}{c}\gamma_{u} & 0 & 1 & \gamma_{u}\end{array}\right]\label{eq:72}\end{equation}
then

\begin{equation}
A^{\mu}=\mathcal{L}_{\nu}^{\mu}\bar{A}^{\nu}\label{eq:73}\end{equation}
Since the molecule's velocity, as measured in $\bar{S}$, is $\bar{v}^{\nu}=K^{\nu}$
we have \begin{equation}
v^{\mu}=\mathcal{L}_{\nu}^{\mu}K^{\nu}=\gamma_{k}\left(\gamma_{u}\left(u+k_{1}\right),k_{2},k_{3},\gamma_{u}c\left(1+\frac{uk_{1}}{c^{2}}\right)\right)\label{eq:74}\end{equation}

In order to show the invariance of the volume element $dv^{*}=cd^{3}v/v^{4}$
we start from 

\begin{equation}
d^{3}v=Jd^{3}K\label{eq:75}\end{equation}
where the Jacobian is given by\begin{equation}
J=det\left[\frac{\partial v^{a}}{\partial K^{b}}\right]\label{eq:76}\end{equation}
and is calculated as follows\[
\frac{\partial v^{a}}{\partial K^{b}}=\frac{\partial}{\partial K^{b}}\left[\mathcal{L}_{\nu}^{a}K^{\nu}\right]=\mathcal{L}_{\nu}^{a}\frac{\partial}{\partial K^{b}}\left[K^{\nu}\right]=\begin{cases}
\gamma_{u}\left(\delta_{b}^{1}-\frac{u}{c}\frac{K_{1}}{K_{4}}\right) & \qquad a=1\\
\delta_{b}^{a} & \qquad a\neq1\end{cases}\]
where use has been made of the fact that, since $K^{\mu}$ is a four-velocity,
$K^{\mu}K_{\mu}=-c^{2}$ and thus\begin{equation}
0=K_{\mu}\frac{\partial K^{\mu}}{\partial K^{b}}=K_{4}\frac{\partial K^{4}}{\partial K^{b}}+K_{a}\frac{\partial K^{a}}{\partial K^{b}}=K_{4}\frac{\partial K^{4}}{\partial K^{b}}+K_{b}\label{eq:77}\end{equation}
Then, the Jacobian is\begin{equation}
J=\gamma_{u}\left(1-\frac{u}{c}\frac{K_{1}}{K_{4}}\right)=\frac{1}{K_{4}}\gamma_{u}\left(K_{4}-\frac{u}{c}K_{1}\right)=\frac{1}{K^{4}}\gamma_{u}\gamma_{k}\left(c+\frac{u}{c}k_{1}\right)=v^{4}\label{eq:78}\end{equation}
and thus\begin{equation}
\frac{d^{3}v}{v^{4}}=\frac{d^{3}K}{K^{4}}\label{eq:79}\end{equation}

Regarding the scalar product $\mathcal{U}^{\nu}v_{\nu}$ we have\begin{equation}
\mathcal{U}_{\nu}v^{\nu}=\mathcal{U}_{\nu}\mathcal{L}_{\mu}^{\nu}K^{\mu}\label{eq:80}\end{equation}
which can be readily calculated using that\begin{equation}
\mathcal{U}_{\nu}=\gamma_{u}\left(u,0,0,-c\right)\label{eq:81}\end{equation}
and the transformation given in Eq. (\ref{eq:72}) as follows\begin{equation}
\mathcal{U}_{\nu}v^{\nu}=\gamma_{u}u\mathcal{L}_{\mu}^{1}K^{\mu}-c\gamma_{u}\mathcal{L}_{\mu}^{4}K^{\mu}=\gamma_{u}^{2}\left[uK^{1}+\frac{u^{2}}{c}K^{4}-uK^{1}-cK^{4}\right]=\gamma_{u}^{2}\left[\frac{u^{2}}{c^{2}}-1\right]cK^{4}\label{eq:82}\end{equation}
and thus\begin{equation}
\mathcal{U}_{\nu}v^{\nu}=-cK^{4}\label{eq:83}\end{equation}
The results in Eqs. (\ref{eq:79}) and (\ref{eq:83}) allow the calculation
of moments of the distribution function in terms of the chaotic velocity
in as similar way as in the non-relativistic case:\begin{equation}
\int\exp\left(\frac{\mathcal{U}^{\beta}v_{\beta}}{zc^{2}}\right)\mathcal{J}dv^{*}=\int\exp\left(-\frac{\gamma_{k}}{c}\right)\mathcal{J}dK^{*}\label{eq:84}\end{equation}
where $\mathcal{J}$ is an arbitrary tensor.

Now we turn to the proof of Eqs. (\ref{eq:47}) and (\ref{eq:48}).
Firstly, since $\mathcal{L}_{4}^{\alpha}=\frac{\mathcal{U}^{\alpha}}{c}$,
\begin{equation}
\mathcal{R}_{4}^{\mu}=h_{\alpha}^{\mu}\mathcal{L}_{4}^{\alpha}=h_{\alpha}^{\mu}\frac{\mathcal{U}^{\alpha}}{c}=0\label{eq:85}\end{equation}
To obtain Eq. (\ref{eq:48}), we separate two cases. For $\mu=4$,
since $h_{4}^{4}=1-\gamma^{2}$ and $h_{b}^{4}=\gamma\frac{\mathcal{U}_{b}}{c}$
\begin{equation}
\mathcal{R}_{a}^{4}=h_{\alpha}^{4}\mathcal{L}_{a}^{\alpha}=\left(1-\gamma^{2}\right)\frac{\mathcal{U}_{a}}{c}+\gamma\frac{\mathcal{U}_{b}}{c}\mathcal{L}_{a}^{b}\label{eq:86}\end{equation}
For the second term we use that\begin{equation}
\frac{\mathcal{U}_{b}}{c}\mathcal{L}_{a}^{b}=\frac{\mathcal{U}_{b}}{c}\left(\delta_{a}^{b}+\frac{\mathcal{U}_{a}\mathcal{U}^{b}}{c^{2}\left(\gamma+1\right)}\right)=\gamma\frac{\mathcal{U}_{a}}{c}\label{eq:87}\end{equation}
and thus\begin{equation}
\mathcal{R}_{a}^{4}=\frac{\mathcal{U}_{a}}{c}=\mathcal{L}_{a}^{4}\label{eq:88}\end{equation}
Finally, for $\mu=\ell=1,2,3$\begin{equation}
\mathcal{R}_{a}^{\ell}=h_{\alpha}^{\ell}\mathcal{L}_{a}^{\alpha}=h_{b}^{\ell}\mathcal{L}_{a}^{b}+h_{4}^{\ell}\mathcal{L}_{a}^{4}\label{eq:89}\end{equation}
or, using that $h_{4}^{\ell}=-\gamma\frac{\mathcal{U}^{\ell}}{c}$
\begin{equation}
\mathcal{R}_{a}^{\ell}=\left(\delta_{b}^{\ell}+\frac{\mathcal{U}_{b}\mathcal{U}^{\ell}}{c^{2}}\right)\mathcal{L}_{a}^{b}-\gamma\frac{\mathcal{U}^{\ell}\mathcal{U}_{a}}{c^{2}}\label{eq:90}\end{equation}
Now, by introducing Eq. (\ref{eq:87}) in Eq. (\ref{eq:90}) one obtains\begin{equation}
\mathcal{R}_{a}^{\ell}=\mathcal{L}_{a}^{\ell}\label{eq:91}\end{equation}
This completes the proof.
\selectlanguage{spanish}%

\end{document}